\documentclass{osa-article}

\journal{oe}

\articletype{Research Article}

 \usepackage[normalem]{ulem}
 
\begin{document}

\title{Fully--correlated multi--mode pumping for low--noise dual--frequency VECSELs}

\author{Gr\'egory Gredat,\authormark{1} Debanuj Chatterjee,\authormark{1}  Ghaya Baili,\authormark{2}, Fran\c{c}ois Gutty,\authormark{2} Isabelle Sagnes,\authormark{3} Fabienne Goldfarb,\authormark{1} Fabien Bretenaker,\authormark{1,4,5} and Hui Liu\authormark{1, *}}

\address{\authormark{1} Laboratoire Aim\'e Cotton, CNRS, Universit\'e Paris--Sud,
ENS Paris--Saclay, Universit\'e Paris--Saclay, Orsay, France.\\
\authormark{2} Thales Research \& Technology, Palaiseau, France.\\
\authormark{3} Centre de Nanosciences et Nanotechnologie (C2N), CNRS, Universit\'e Paris--Sud, Universit\'e Paris--Saclay, Marcoussis, France. \\
\authormark{4} Light and Matter Physics Group, Raman Research Institute, Bangalore 560080, India. 
}

\email{\authormark{5} fabien.bretenaker@u-psud.fr\\
\authormark{*} hui.liu@u-psud.fr} 


\begin{abstract}
We report a fully--correlated multi--mode pumping architecture optimized for dramatic noise reduction of a class--A dual--frequency Vertical External Cavity Surface Emitting Laser (VECSEL). Thanks to amplitude division of a laser diode, the two orthogonally polarized modes emitted by the VECSEL oscillating at 852 nm are separately pumped by two beams exhibiting fully in--phase correlated intensity noises. This is shown to lead to very strong and in--phase correlations between the two lasing modes intensities. As a result, the phase noise power spectral density of the RF beat note generated by the two modes undergoes a drastic reduction of about 10 to 20 dB throughout the whole frequency range from 10 kHz to 20 MHz and falls below the detection floor above a few MHz. A good agreement is found with a model which uses the framework of rate equations coupled by cross--saturation. The remaining phase noise is attributed to thermal effects and additional technical noises and lies mainly within the bandwidth of a phase--locked--loop.
\end{abstract}


\section{Introduction}
Microwave photonics, based on low noise tunable optically--carried microwave signals, addresses many applications such as frequency standards distribution \cite{Ma:94}, photonic remoting for antenna in radar systems, photonic links for cellular, wireless, satellite and radio--astronomy applications, cable television systems, optical signal processing  \cite{capmany:07}, etc. In contrast with other technologies, dual--frequency lasers can provide, thanks to their beat note, optically carried RF signals and offer the advantage of optimal modulation depth \cite{Alouini:2001}. The two laser modes share the same cavity, their phase noises are thus expected to cancel in the beat note. To get a tunable RF signal with such a dual--frequency laser, a birefringent crystal (BC) can be inserted inside the cavity to create two optical paths for two orthogonal polarization modes. Besides, VECSELs exhibit class--A dynamics \cite{Baili:09} since the photon lifetime inside the cavity can be much longer than the carriers lifetime inside the quantum wells. Free from relaxation oscillations and displaying low noise, dual--frequency VECSELs are therefore good candidates for microwave photonics. Their intensity and phase noises have already been studied in details at 1~$\mu$m \cite{De:14} and 1.55~$\mu$m wavelengths, and more recently at 852 nm \cite{Liu:18}.

 The investigation of this latter wavelength has been triggered by potential new applications such as cesium atomic clocks \cite{Dumont:14} or sensors based on coherent population trapping (CPT). For instance, a double lambda scheme for CPT probed by $\mathrm{lin}\perp\mathrm{lin}$ laser beams is shown to create Raman--Ramsey fringes with a larger contrast than the usual simple lambda scheme in \cite{Zanon:2005}.
 
  All these potential applications of dual--frequency VECSELs at various wavelenths require low intensity and beatnote phase noises. A deep understanding of the noise mechanisms is the cornerstone of our approach reported in \cite{Liu:18} and a key point for noise reduction strategies. In particular, intensity noise reduction and phase noise reduction are shown to rely on the same four rules :

(i) the pump noise should be obviously made as small as possible;

(ii) the excitation ratios for the two lasing modes have to be balanced;

(iii) the cross--saturation between the two modes must be minimized and;

(iv) the correlations between the intensity noises of the pumping regions corresponding to each mode have to be in--phase and as strong as possible.

 On the one hand, minimizing the cross--saturation can be achieved by increasing the spatial separation between the two modes in the active medium.
On the other hand, in--phase 100\% correlation between the two pumping areas can be achieved trivially using a single--mode fibered pump (provided it delivers enough power) \cite{Liu:18OL} instead of a multi--mode fibered one that would create a speckle pattern on the VECSEL structure. Indeed, single--mode fibered pumping allows the gain areas for the two modes areas in the gain medium to intercept the same noise. However, this solution cannot be applied to several wavelengths for which either there simply exists no commercial single--mode fibered pump laser, or there is a power issue. Pumping 1.5 $\mu$m dual--frequency VECSELs using currently available single-mode fibered pump diodes requires a trick, as reported in \cite{Liu:18OL}, otherwise the power delivered to the structure is not sufficient. In the particular case of 852 nm VECSELs for example, no commercial single--mode fibered laser diode is currently available with enough power. Yet another way to directly increase the pump noises correlation is to superimpose as much as possible the two pumping regions so that they can intercept the same intensity fluctuations. Hence, fulfilling both conditions (iii) and (iv) simultaneously seems contradictory. The question then is to find a way to get low cross--saturation and strong in--phase correlations between the pumping regions at the same time using a multi--mode fibered pump. 

In this paper, we address this issue for noise reduction in a dual--frequency VECSEL at 852 nm using a new pump architecture. Using amplitude division of a laser diode beam, the two lasing modes are separately pumped. Fitting the modes with these two copies of the same pump and minimizing their overlap are key points to achieve the intended goal of phase noise improvements with multi--mode pumping.

The paper is organized as follows. The dual--frequency VECSEL and its novel pumping scheme are presented in Section \ref{section1}. A particular attention is paid to the pump noise features, the correlation between the two pumping regions and the cross--saturation level. 
 Section \ref{section2} focuses on the intensity noise of each mode and the noise mechanisms driven by our pumping set--up. Section \ref{section3} is dedicated to the analysis of the phase noise of the beat note. We perform a comparison with the previous studies \cite{Liu:18, Dumont:14}, which use a single pump spot on the gain medium. Experimental results are also compared to the model developed in \cite{Liu:18}. Finally, we sum up our results in Section \ref{conclusion}.

\section{Device implementation for fully--correlated pumping} \label{section1}

\begin{figure}[!ht]
\centering\includegraphics[width=0.75\textwidth]{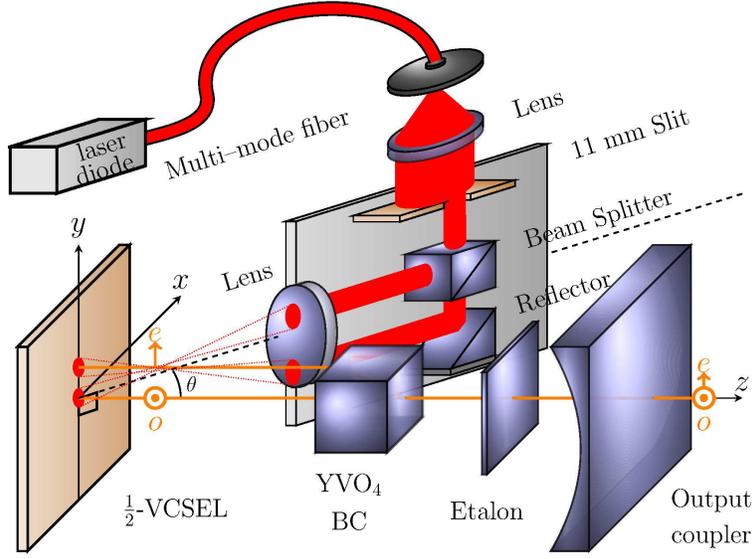}
\caption{Experimental set-up for fully in-phase correlated pumping. The cross-polarized ordinary (o) wave and extraordinary (e) wave created by the birefringent crystal (BC) are separately pumped thanks to amplitude division of a multi--mode fibered laser diode.}
\label{fig1}
\end{figure}

The dual--frequency VECSEL and its pumping architecture are schematized in Fig. \ref{fig1}. The semiconductor chip (referred to as 1/2--VCSEL) is glued to a Peltier cooler, which is bonded to a heat sink. The Peltier temperature is stabilized at 20$^{\circ}$C. This 1/2--VCSEL is a multi-layered structure grown on a 350-$\mu$m-thick GaAs substrate by metal-organic chemical-vapor deposition method. It contains a distributed Bragg reflector and active layers. The Bragg reflector is composed of 32.5 pairs of AlAs/Al$_{0.22}$GaAs quarter-wave layers leading to a reflectivity larger than 99.94\% around 850~nm. This mirror constitutes a half of the cavity formed with the output coupler which is a concave mirror with a transmission of 0.5\% and a radius of curvature of 5~cm. 
In the active layers, seven 8-nm-thick GaAs quantum wells are embedded in Al$_{0.22}$GaAs barriers, which absorb the pump power. About 75\% of the pump power at 673~nm can be absorbed in a single pass. Each of the quantum wells is located at an antinode of the laser field. Two layers of Al$_{0.39}$GaAs form potential barriers to confine the carriers. A 50-nm-thick InGaP and 5~nm GaAs layer cap the structure to prevent the Al oxidation. This chip is designed without anti-reflection coating to increase the gain of the mode resonant within the micro-cavity created in the semiconductor structure. If we consider the losses of the output coupler, the photon lifetime inside the empty 5-cm-long cavity is 32~ns, which is much longer than the carrier lifetime inside the wells $\tau~\simeq~1~\mathrm{ns}$, thus ensuring that the laser obeys class-A dynamics \cite{Baili:09}.

An anti-reflection coated YVO$_4$ BC is inserted inside the cavity. It is cut at 45$^{\circ}$ off its optical axis, thus creating a walk--off between the extraordinary polarization and the ordinary one. This separation leads to the oscillation of two orthogonally $x$- and $y$-polarized laser modes inside the cavity simultaneously. However, depending on the mode radius $w_0$ in the structure, the two modes partially overlap inside the gain structure and thus experience some competition through nonlinear coupling, i.e, gain cross-saturation. This is described by the self- to cross- saturation ratios denoted as $\xi_{xy}$ and $\xi_{yx}$ and the coupling constant $C = \xi_{xy} \cdot \xi_{yx}$. Longitudinal single-mode operation is obtained for each of the two orthogonal polarizations by inserting an isotropic YAG etalon. The 100 $\mu$m--thickness of the etalon leads to a beat note frequency in the range of few hundreds of MHz.

  In order to minimize the cross--saturation (condition (iii)), the walk--off can be increased by increasing the thickness of the BC. But this would lead to an increase of the intra--cavity losses. The thickness of the BC chosen here results from a trade--off between these two effects and is equal to  1~mm, leading to a value of the walk--off equal to 100~$\mu$m. With this crystal and a cavity length equal to $L_\mathrm{cav} = 46.8\,\mathrm{mm}$, the two modes are almost completely separated in the active medium and we estimate the coupling constant to be as low as $C \simeq 0.05$, which is three times smaller than the value obtained in \cite{Liu:18}.
 This reduced value of the coupling constant is promising for phase noise reduction provided condition (iv) is simultaneously fulfilled.

 The pump laser is a 635~nm laser diode. It is delivered to a connector by a multi--mode fiber whose core diameter is  $102$~$\mu\mathrm{m}$  with a numerical aperture equal to 0.22. 
 Since the separating distance between the two modes is about  $100$~$\mu\mathrm{m}$, which is larger than the mode radius $w_0$ in the 1/2--VCSEL, pumping the gain medium with a single spot would lead the two modes pumping regions to intercept poorly correlated intensity fluctuations from the speckle pattern. This is not compatible with condition (iv) and is thus a dead--end for phase noise reduction. As a solution to make the two spatially separated modes undergo the same pump intensity fluctuations, we propose to use two spatially separated pump beams originating from the same source. 
The diode laser beam is collimated with a lens and shaped with a slit. About two thirds of the power are transmitted. The laser beam is then sent to a beam-splitter which performs an amplitude division and thus creates two copies of the same pump. A second lens is then used to image the pumps onto the semiconductor chip with an incidence angle $\theta  \simeq 40^{\circ}$ and a working distance of 30 mm. As a result, two approximatively $100 \,\mu\mathrm{m} \times 70 \,\mu\mathrm{m}$ elliptical spots are able to pump each mode on the structure.
 Thanks to the $70 \,\mu\mathrm{m}$ typical width of the pump spot along $y$ direction, a walk--off of $100$~$\mu\mathrm{m}$ indeed enables to pump separately the two modes regions without overlap, meaning that each beam pumps one and only one mode. However, this is true only if the two pump spots fit the separating distance between the two modes. To this aim, the reflector used after the beam-splitter is put on an adjustable stage and can therefore be tilted. The pumps are then responsible for the population inversion of each mode through the creation of reservoirs of carriers whose average unsaturated numbers are denoted as $\overline{N}_{0i}$ for $i = \left\lbrace x, y\right\rbrace$ and their fluctuations $\delta N_{0i}$.
  As a consequence, each pump, whose effective power is denoted as $P_{\mathrm{p},i}$, enables its own laser operation with the excitation ratio noted $r_i$.
 Due to the choice of amplitude division to generate these two pumps, full correlation of the intensity noises of the two beams is expected. The correlation amplitude is denoted as $\eta$ while its phase is denoted $\Psi$. With $\mathrm{RIN_p}$ the Relative Intensity Noise of the pump, the correlation between the modes verifies :

  \begin{equation}
  \label{eq1}
   \left\langle \widetilde{\delta N}_{0x}\left(f\right) \cdot \widetilde{\delta N}^{\ast}_{0y} \left(f\right)\right\rangle = \eta\,\mathrm{RIN_p}\left(f\right)\,\overline{N}_{0x}\,\overline{N}_{0y}\,e^{i\,\Psi} \, ,
  \end{equation}
  where $\left\langle \cdot \right\rangle$ stands for the statistical average, tilde denotes the Fourier transformed quantities and $f$ the noise frequency.
  
   \begin{figure}[!ht] 
\centering \includegraphics[width=.49\textwidth]{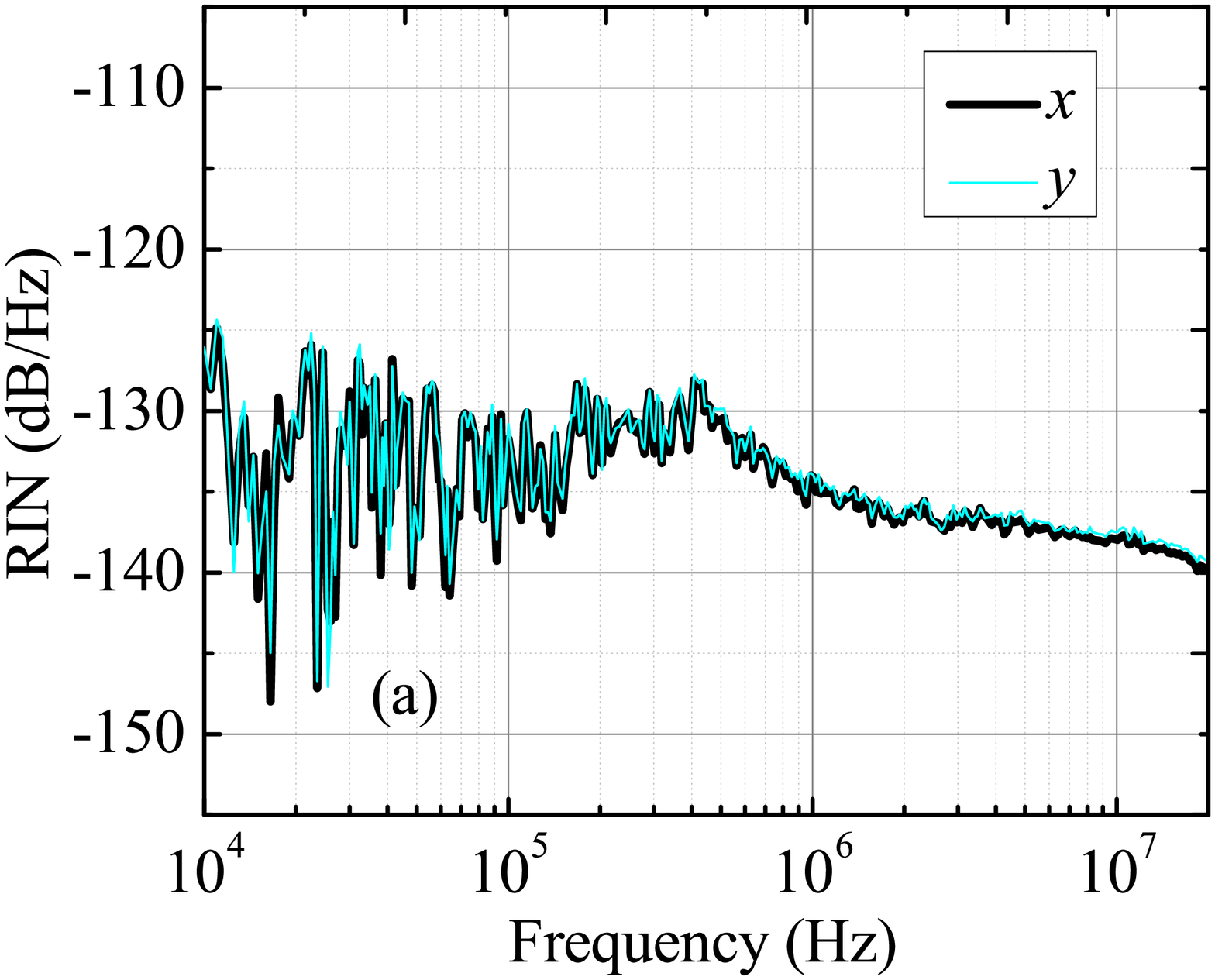} \includegraphics[width=.49\textwidth]{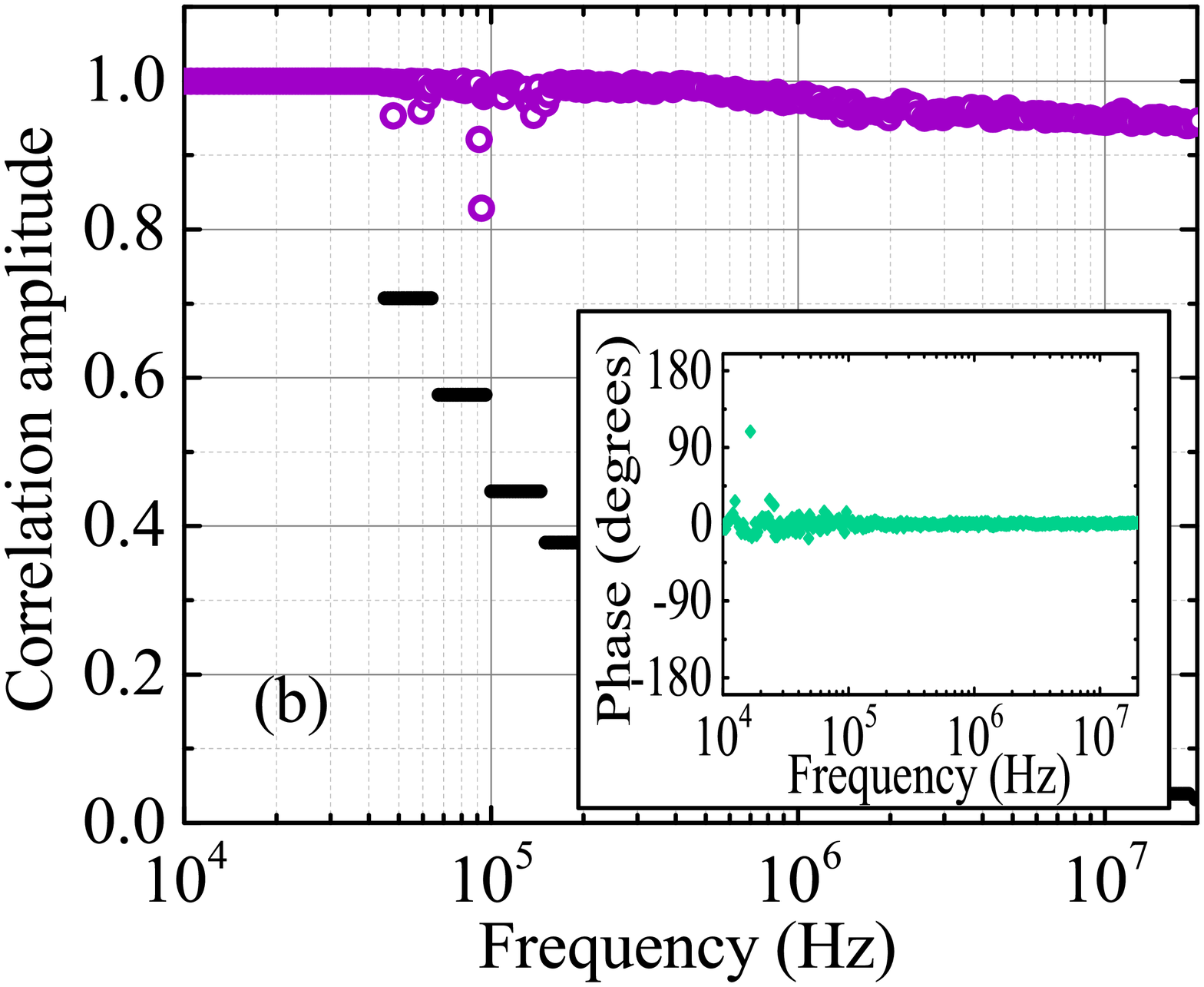}
\caption{Pump noise measurements. The thick (black) line in panel (a) shows the RIN spectrum of the beam pumping $x$-polarization while the thin (cyan) line stands for $y$-polarization. Panel (b) shows the correlation spectra between the two pump beams. The correlation amplitude is represented by open circles (in purple), the detection floor by thick lines (in black) and the correlation phase lies inside the devoted inset (green diamonds).}
\label{fig2}
\end{figure}

  Figure \ref{fig2} focuses on the experimental noise properties of the two pumps, namely the pump RIN level and the correlation between the RINs of the two pumps. The RIN spectra of the pump beams are displayed in Fig.~\ref{fig2}(a). They can be modeled by a constant value $\mathrm{RIN_p} = - 133 \,\mathrm{dB/Hz}$. Figure \ref{fig2}(b) evidences a very high correlation amplitude, very close to 1 at low frequencies and around 0.95 after a few MHz. This figure also highlights a fully in--phase behavior of the pumps since $\Psi =0$. 
  
  The pumping scheme shown in Fig.~\ref{fig1} is thus able to overcome the apparent contradiction between conditions (iii) and (iv). Indeed, very strongly in--phase correlated pumps are driving the dual--frequency VECSEL while very low cross--saturation is displayed.
  
  In the following, the pump noise correlation amplitude is modeled by a constant value $\eta = 0.98$. Together with the coupling constant $C = 0.05$, this forms a couple of laser parameters in strong contrast with the values ($\eta=0.45$~,~$C=0.44$) and ($\eta=0.1$~,~$C=0.15$) previously achieved in \cite{Liu:18}.  We can thus expect these low mode coupling and strong in--phase pump noise correlation to have a positive impact on the noises of our dual--frequency VECSEL.

\section{Analysis of the in--phase and anti--phase components of the laser intensity noise} \label{section2}

 \begin{figure}[!ht] 
\centering\includegraphics[width=0.49\textwidth]{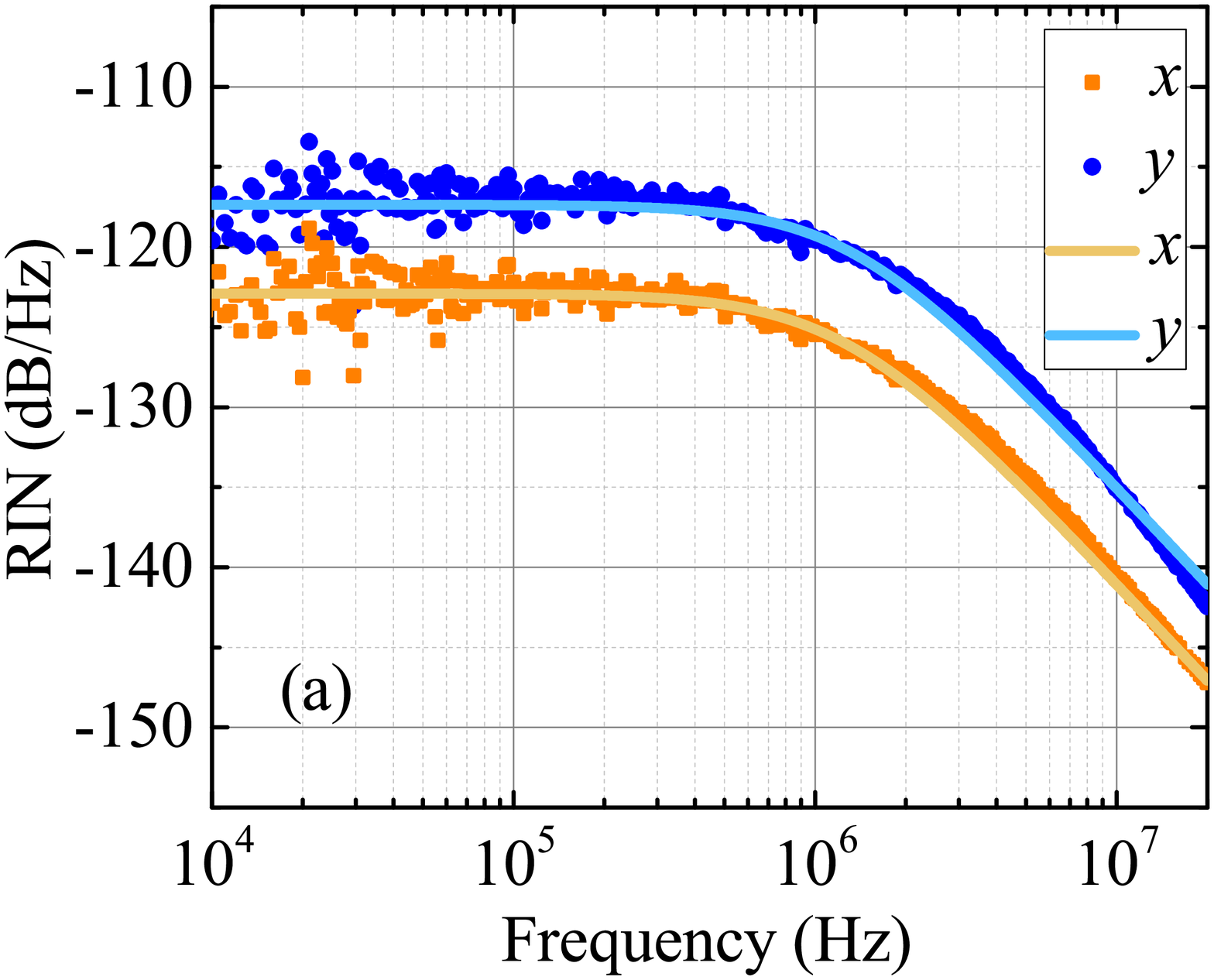} \includegraphics[width=0.49\textwidth]{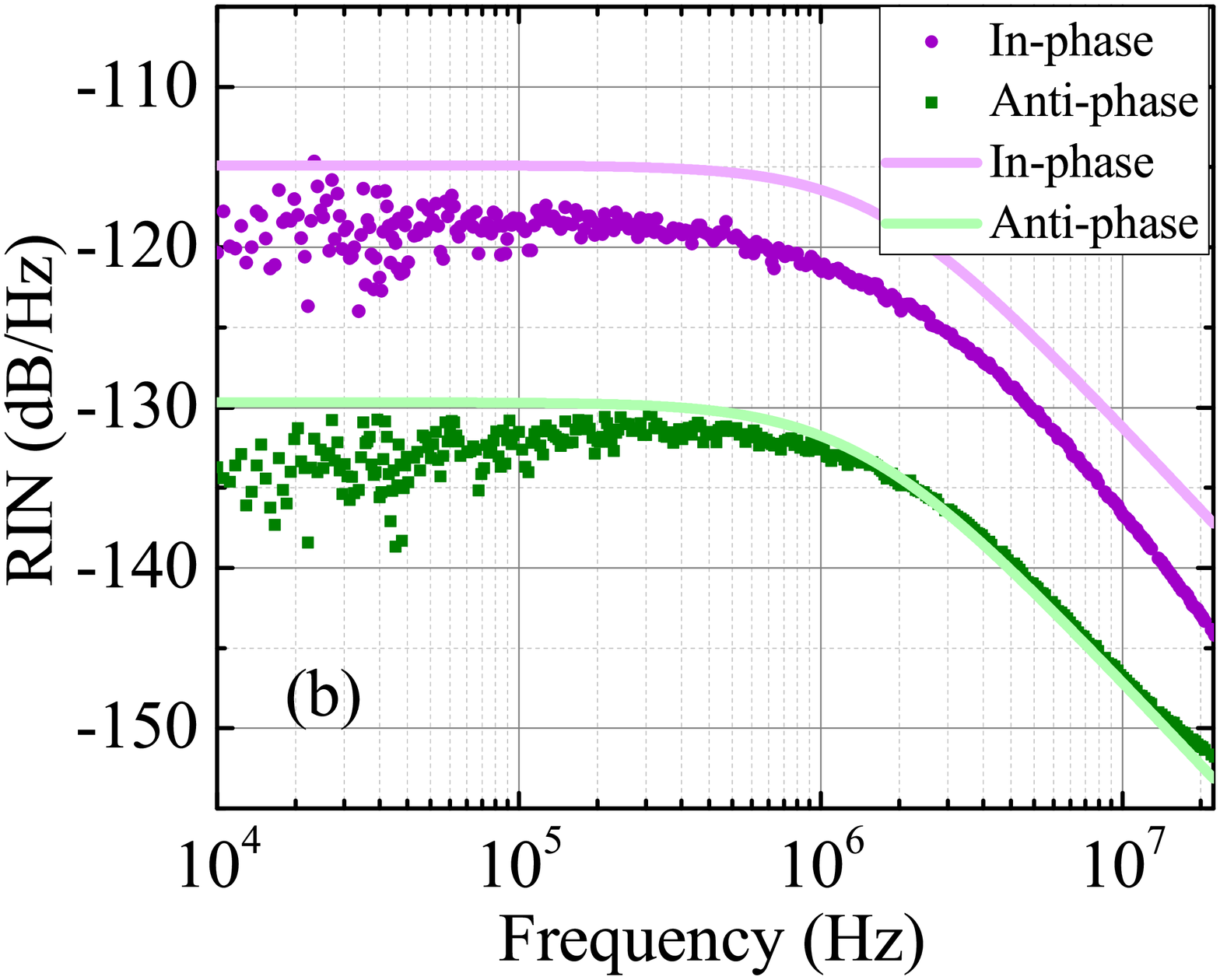}
\caption{Intensity noise spectra of the dual--frequency VECSEL. The symbols stand for measurements and the solid lines stand for the model. Panel (a) shows the RIN of $x$--polarization (in blue) and the RIN of $y$--polarization (in orange). Panel (b) shows the RIN of the in-phase noise mechanism (in purple) and the RIN of the anti-phase noise (in green). The parameters used for the model are : $\tau = 1\,\mathrm{ns}$, $\Psi=0$, $\eta =0.98$, $r_x=1.38$, $r_y=1.23$, $C=0.05$, $\mathrm{RIN_p}=-133\,\mathrm{dB/Hz}$, $\tau_x=30\,\mathrm{ns}$, $\tau_y = 17\,\mathrm{ns}$.}
 \label{fig3}
\end{figure}

Using a half--wave plate then a polarization beam splitter, the  $x$- and $y$-polarized modes emitted by the dual--frequency VECSEL are separately sent to two photodiodes. After amplification, the intensity fluctuations are finally recorded by an oscilloscope and analyzed.  Figure~\ref{fig3}(a) shows that the RINs exhibit the usual behavior of a low--pass filter whose cut--off frequency is typically  given by $1/\tau_i$, where $\tau_i$ corresponds to the photon lifetime inside the cavity for polarization $i = \left\lbrace x, y\right\rbrace$. Figure 3 compares these measurements to a model which takes into account the influence of the pump noise on the RIN through coupled rate equations \cite{Liu:18, De:14} theory. A very good agreement is found. The eigenmodes of the intensity noises in dual--frequency oscillation can be interpreted in terms of in--phase and anti--phase modes just like for coupled pendulums. Figure \ref{fig3}(b) evidences the fact that the anti--phase noise is between 8 and 15 dB lower than the in--phase noise throughout the whole frequency range 10 kHz -- 20 MHz. This confirms the expectation that the full in--phase pumping combined with a very low non--linear coupling constant $C $  are favorable conditions for the in--phase noise to dominate. Indeed, such a domination of the in--phase mechanism was predicted in\cite{Liu:18}, which states that decreasing the coupling between the two modes favors in--phase noise with respect to anti--phase noise.  The dramatic anti--phase noise reduction corroborates the fact that the gain competition between the two modes is well inhibited. This proves that the present laser architecture prevents the noises of the dual--frequency VECSEL from being deteriorated by cross--saturation. 

 \begin{figure}[!ht]  
 \centering\includegraphics[width=.49\textwidth]{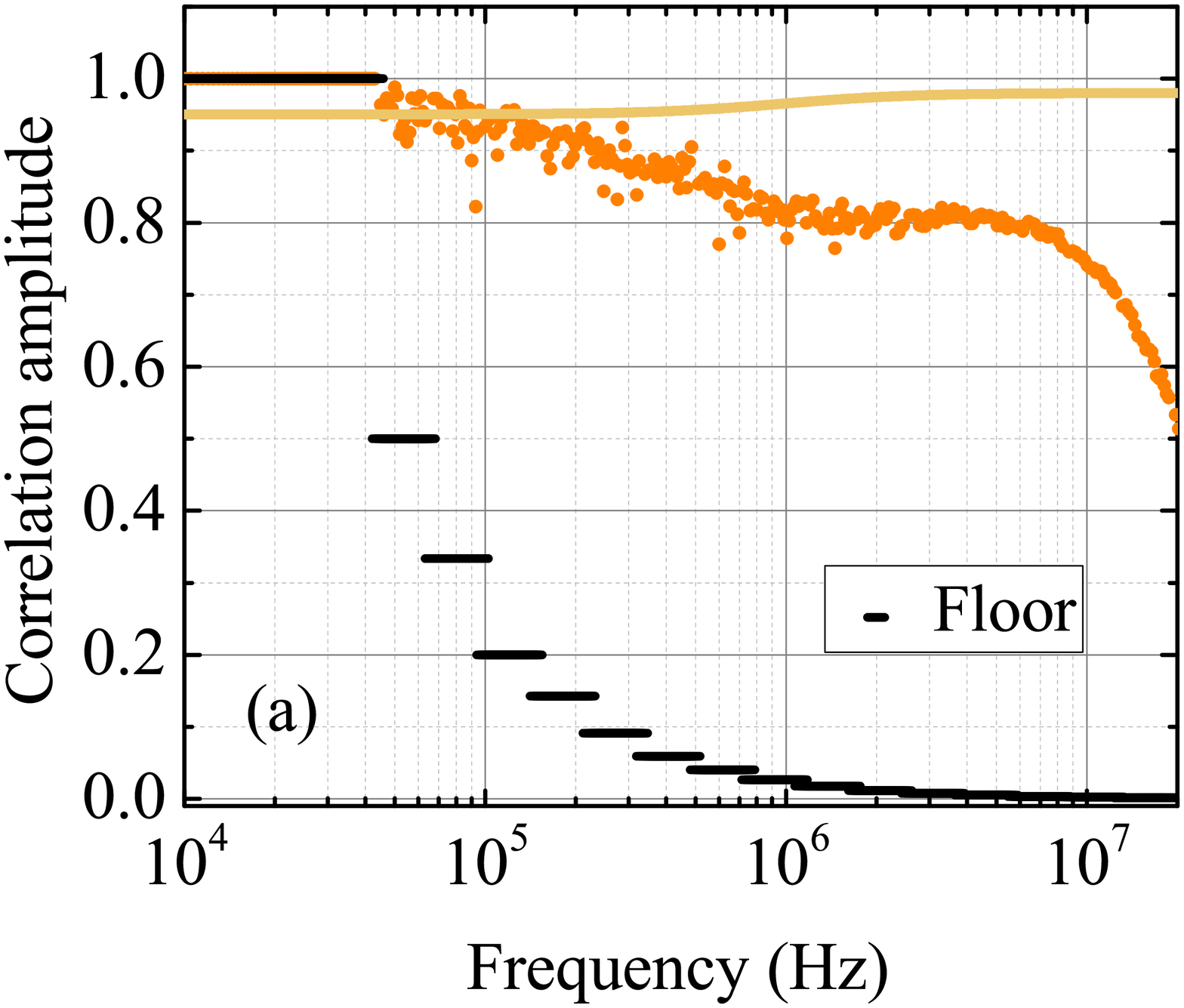} \includegraphics[width=.49\textwidth]{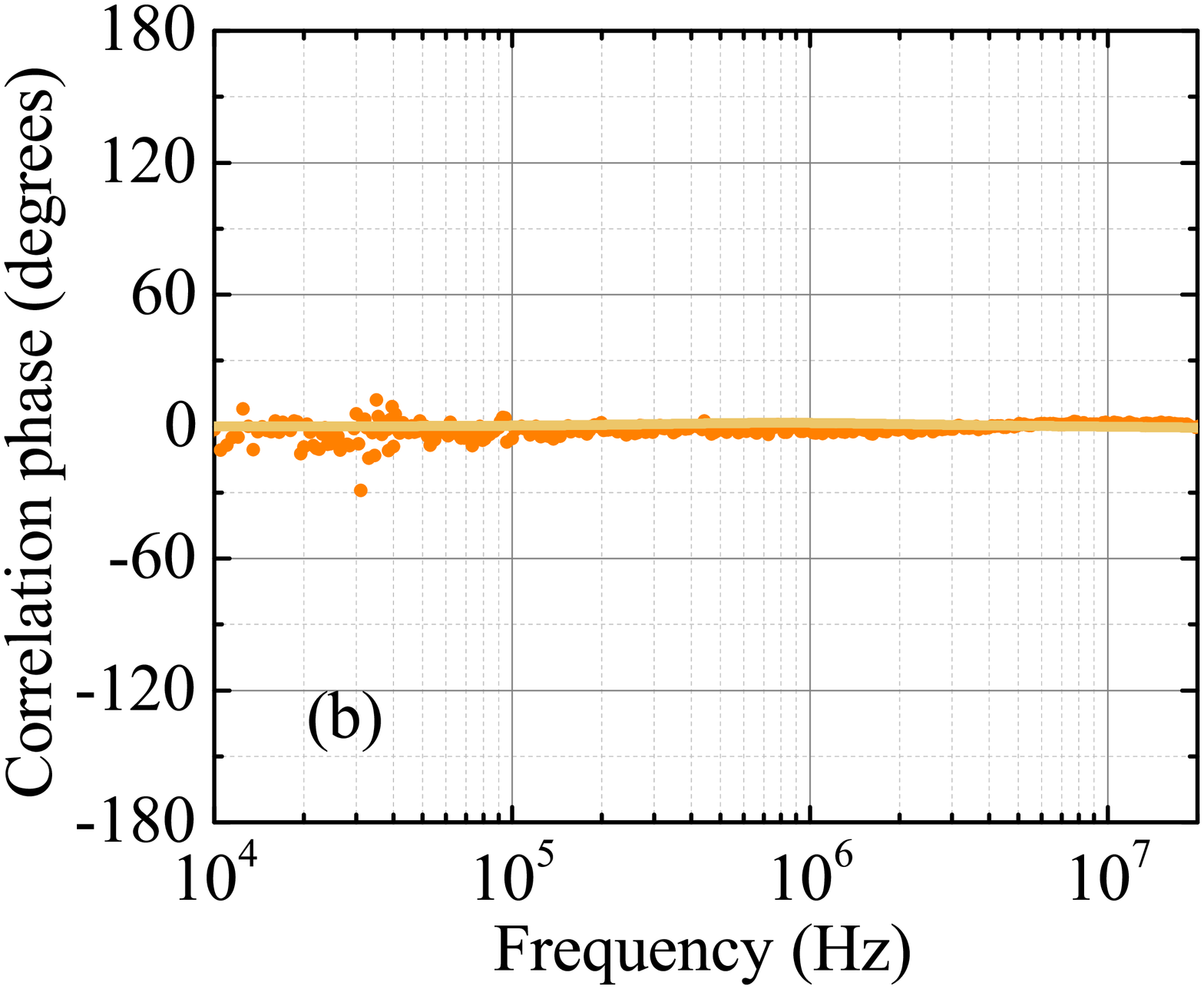}
\caption{Amplitude (a) and phase (b) of the correlation spectrum between the intensity noises of the two modes. The symbols stand for measurements and the solid lines stand for the model computed with the same parameters as in Fig.~\ref{fig3}.}
\label{fig4}
\end{figure}

The amplitude and phase of the resulting correlation of the dual--frequency VECSEL intensity noises are analysed in Fig.~\ref{fig4}. The two modes intensity noises are found to be  fully in--phase correlated with a strong correlation amplitude. More precisely, Fig.~\ref{fig4}(a) shows that the correlation amplitude between the two modes is larger than 0.8 up to several MHz. The decrease at frequencies larger than 8 MHz is due to the limitations of our measurement. Indeed, at this frequency the intensity noise approaches -140 dB/Hz (as shown in Fig.~\ref{fig3}), which is very close to the shot--noise level. This explains why this decrease in correlation amplitude is not predicted by the model.  Figure~\ref{fig4}(b)  corroborates the fact that in--phase noise mechanism is strongly dominant between 10 kHz  and  20 MHz. 

To summarize, we have proved in this section that, when conditions (iii) and (iv) are both verified, a significant reduction of the dual--frequency VECSEL anti--phase intensity noise is observed, as expected from our modelling. The next section aims at investigating whether a similar decrease in the phase noise of the beat note is obtained.

\section{Phase noise reduction} \label{section3}

First, it is worth mentioning some phase noise properties of VECSELs. These semiconductor lasers exhibit large Henry $\alpha$ factors responsible for the coupling between the phase and the amplitude fluctuations of the laser field \cite{Henry:1982}.  This effect of the Henry factor has been shown to be the dominating mechanism for the phase noise of the beat note of dual--frequency VECSELs at high frequencies \cite{De:14}, typically above one MHz.   Another relevant contribution to the phase noise has been shown to originate from the thermal fluctuations of the semiconductor active medium induced by pump intensity fluctuations. This thermal effect dominates the beat note phase noise at low frequencies and its properties can be described using a macroscopic model which involves three parameters \cite{Laurain:10} : $R_\mathrm{T}$, the thermal resistance of the semiconductor structure ; $\tau_\mathrm{T}$, its thermal response time ; and $\Gamma_\mathrm{T}$, its refractive index variation with temperature. 

Besides, for dual--frequency VECSELs, the non--linear coupling between the modes, induced by cross--saturation, has a detrimental influence on the phase--noise \cite{Liu:18}. That is why, in the introduction, condition (iii)  stipulates a very low coupling constant to achieve a low phase noise. Here, we have estimated  the coupling constant to be only $C = 0.05$.
The beat note phase--noise level is also expected to strongly depend on the correlations between the pump noises seen by the two modes \cite{De:14}, which thus deserve to be optimized.  Condition (iv) stipulates that fully in--phase correlated pumping  is required, i.e. $\eta \to 1$ and $\Psi \to 0$.  With the present specially designed pumping architecture, Section~\ref{section2} has shown that we achieve a correlation amplitude $\eta$ close to 1 with $\Psi =0$.
In order to also meet condition (ii),  balanced excitation ratios ($r_x \simeq r_y$) are needed. This latter condition plays though a marginal role to reduce the phase noise as soon as $\eta \neq 1$, which is the case here.
   
\begin{figure}[!ht]  
 \centering\includegraphics[width=7.5cm]{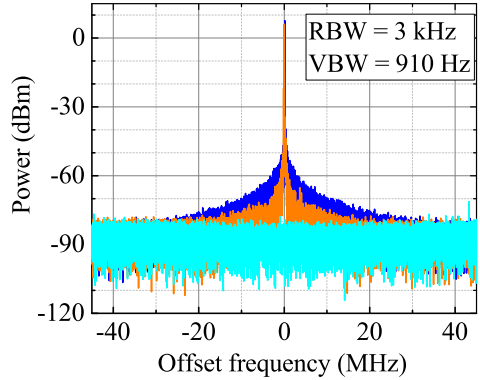}
\caption{Beat note spectrum obtained with an electrical spectrum analyzer. The bottom light trace (in cyan) stands for the detection floor. The beat note spectra are plotted versus the frequency offset from the carrier. The wider spectrum (in blue) corresponds to the single pump spot scheme reported in \cite{Liu:18} whereas the other one (in orange) corresponds to the two fully in--phase correlated pumps.}
\label{fig5}
\end{figure}

Figure \ref{fig5} reports the VECSEL output beat note detected with a spectrum analyser in two respective conditions : with a single pump spot as reported in \cite{Liu:18} (in blue) and with the two in--phase correlated pumps (in orange). As a consequence of the novel pumping scheme and as predicted, the beat note noise pedestal experiences a tremendous decrease.

\begin{figure}[!ht] 
\centering\includegraphics[width=7.5cm]{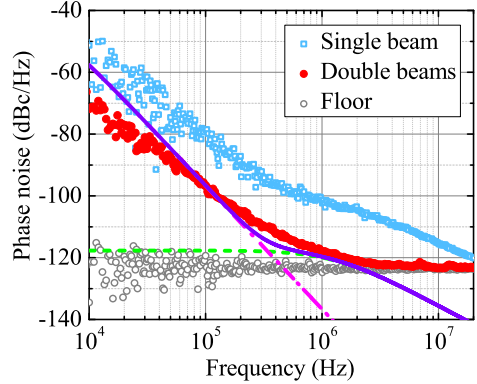}
\caption{Power spectral density spectra of the beat note phase noise. The symbols are experimental measurements. The dots (in red) correspond to the pumping scheme with two beams fully in--phase correlated while the open squares (in light--blue) stand for the single pump--spot scheme reported in \cite{Liu:18}. The open gray circles represent the detection floor. The total phase noise model with the two pumps is plotted with a solid line (in dark purple). The dashed line (in green) is the contribution of the phase-amplitude coupling. The dash--dotted line (in magenta) is the contribution of the thermal effects. They are computed with $\alpha = 5.2$, $P_{\mathrm{p},x} = 0.48 \,\mathrm{W}$, $P_{\mathrm{p},y} = 0.45\,\mathrm{W}$, $R_\mathrm{T} = 40 \,\mathrm{K.W^{-1}}$, $\tau_\mathrm{T} = 30 \,\mu\mathrm{s}$, $\Gamma_\mathrm{T} = 1.39 \times 10^{-7} \,\mathrm{K^{-1}}$ and the same other parameters as in Fig.~\ref{fig3}.}
 \label{fig6}
\end{figure}

 Figure \ref{fig6} plots the power spectral density of phase fluctuations of the VECSEL beat note. This figure evidences a drastic reduction of 10 to 20 dB of the phase noise throughout the whole frequency range from 10 kHz to 20 MHz (red dots) with respect to the former pumping scheme (open light-blue squares). The beat note phase noise falls below the detection floor at about 2 MHz offset from the carrier. Thanks to the model, Fig.~\ref{fig6} also shows that the phase--amplitude coupling contribution to the phase noise power spectral density (dashed green line) has vanished to reach the level of the measurements floor. The remaining part of the phase noise at low frequencies is thus mainly induced by the thermal fluctuations (dash-dotted magenta line) and technical noises. 

It is worth noticing that, in the model for the  thermal noise, the fact that the correlation amplitude $\eta$ of the pump noises is not strictly equal to 1 leads to a larger contribution to the beat note phase noise than the unbalanced pumping of the two modes $P_{\mathrm{p},x} \neq P_{\mathrm{p},y}$. We could thus expect a stronger phase noise suppression to occur with a single--mode fibered pump ensuring $\eta =1$ and delivering enough power like in \cite{Liu:18OL}, provided such a single--mode fiber--coupled high--power diode laser would exist around 673 nm.
Furthermore, we observe a discrepancy reaching almost 10 dB at 10 kHz between the model and the measurements. The experimental spectra exhibit a $f^{-3}$ slope for frequencies below 300 kHz whereas the theoretical model exhibits a larger slope in $f^{-4}$. This means that the model for thermal noise fails reproducing the flicker like frequency noise which is experimentally demonstrated. Therefore, this model would deserve to be further refined. To do so, a microscopic approach would be necessary. Indeed, the theoretical plots in Fig.~\ref{fig6} have been obtained using an oversimplified model in which the frequency response function $\Gamma\left(f \right)$ of the thermal effect is approximated by a simple second--order low--pass filter with a cut--off frequency given by $1/\tau_\mathrm{T}$. $\Gamma\left(f \right)$ could be derived for example using a semi--analytic method like in \cite{Reichling:1994}, which would take into account the detailed structure of the semiconductor chip.

\section{Conclusion} \label{conclusion}
In conclusion, we have demonstrated a tremendous reduction of the beat note phase-noise of a dual--frequency VECSEL operating at 852 nm with multi--mode pumping. This reduction comes as a result of the fully in--phase correlated pumping (with correlation amplitude $\eta = 0.98$ and phase $\Psi=0$) with very low gain--competition ($C=0.05$). This has been made possible using amplitude--division of a multi--mode fibered pump diode, which creates spatially separated pump spot fitting the modes and originating from the same pump source. This work is moreover consistent with the theoretical predictions. This novel multi--mode fibered pump scheme, allowing low noise dual--frequency operation, can be applied to all wavelengths for which there exists no commercial suitable single-mode fibered pump laser diode. Furthermore, even for wavelengths for which single-mode fibered pump lasers exist, this pumping scheme can prove useful to use the larger power of multi--mode pumps while keeping a low phase noise operation.

The dual--frequency VECSEL studied here delivers a beat note in the range of few hundreds of MHz, through the use of an isotropic YAG etalon of thickness 100 $\mu$m. This setup could be adapted to produce higher frequency betnotes by optimizing the etalon, as was recently performed to get a beatnote frequency tunable around 9.2 GHz \cite{Dumont:14}. This makes our setup useful to all possible applications of dual--frequency VECSELs. For example, our noise reduction strategy could be applied to CPT cesium clocks. In this case, the stringent needs in terms of noise for the laser source were evaluated in \cite{Tricot:2018,Dumont:15}. For example, a RIN level as low as -150 dB/Hz up to 1 MHz is required to target a $5\times 10^{-13}$ relative frequency stability at 1 s integration time. Both the use of our novel pumping architecture and implementation of an OPLL could improve the phase and intensity noises. To this aim, a source of low noise RF reference, as reported in \cite{Francois:14}, will be required. The simultaneous stabilization of both the frequency difference on a RF reference and the absolute laser frequency on an atomic transition has also been investigated in \cite{Camargo:13} in such a context. The two frequency lockings were not perturbing one another. Notice also that there should be solutions to reduce the volume and the influence of environmental perturbations on the laser. For example, industrial integration of single-frequency VECSELs is currently being explored \cite{Chomet:18}. Such techniques could represent promising potential solutions for dual--frequency VECSELs in view of developing more compact atomic clocks.

Moreover, the implementation of an OPLL for our dual--frequency VECSEL with improved multi--mode pumping architecture will permit further characterization of the noise properties of these lasers, especially at low frequencies. In particular, a new model for the thermal noise needs to be developed, taking into account the microscopic structure of the semiconductor laser and the particular dual--spot pumping geometry. It will allow us to gain a deeper understanding of the remaining phase noise and will benefit to all applications of dual--frequency VECSELs.

Finally, let us mention that our dual--spot pumping strategy may find applications beyond cw dual--frequency VECSELs. For example, it could be used to reduce the noise of dual--polarization MIXSELs that are presently developed to perform dual--comb spectroscopy \cite{Link:2018}.

\section*{Funding}
 Agence Nationale de la Recherche (grant number ANR-15-CE24-0010-04) ;  Direction G\'en\'erale de l'Armement (DGA). 

\section*{Acknowledgments}
The work of HL, FB, FG, GG and GB is performed in the framework of the joint lab between Laboratoire Aim\'e Cotton and Thales R\&T. The authors thank Gr\'egoire Pillet, Syamsundar De, Christophe Siour, Ga\"elle Lucas-Leclin and Sylvie Janicot for technical help and valuable discussions.


\bibliography{sample}

\begin{thebibliography}{10}
\newcommand{\enquote}[1]{``#1''}

\bibitem{Ma:94}
L.~S. Ma, P.~Jungner, J.~Ye, and J.~L. Hall, \enquote{Delivering the same
  optical frequency at two places: accurate cancellation of phase noise
  introduced by an optical fiber or other time-varying path,}
  {\protect\JournalTitle{Opt. Lett.}} \textbf{19}, 1777--1779 (1994).

\bibitem{capmany:07}
J.~Capmany and D.~Novak, \enquote{Microwave photonics combines two worlds,}
  {\protect\JournalTitle{Nature photonics}} \textbf{1}, 319--330 (2007).

\bibitem{Alouini:2001}
M.~Alouini, B.~Benazet, M.~Vallet, M.~Brunel, P.~D. Bin, F.~Bretenaker, A.~L.
  Floch, and P.~Thony, \enquote{Offset phase locking of {Er,Yb:Glass} laser
  eigenstates for {RF} photonics applications,} {\protect\JournalTitle{IEEE
  Photonics Technology Letters}} \textbf{13}, 367--369 (2001).

\bibitem{Baili:09}
G.~Baili, M.~Alouini, T.~Malherbe, D.~Dolfi, I.~Sagnes, and F.~Bretenaker,
  \enquote{Direct observation of the class-{B} to class-{A} transition in the
  dynamical behavior of a semiconductor laser,}
  {\protect\JournalTitle{Europhysics Letters}} \textbf{87}, 44005--1--44005--5
  (2009).

\bibitem{De:14}
S.~De, A.~E. Amili, I.~Fsaifes, G.~Pillet, G.~Baili, F.~Goldfarb, M.~Alouini,
  I.~Sagnes, and F.~Bretenaker, \enquote{Phase noise of the radio frequency
  ({RF}) beatnote generated by a dual-frequency {VECSEL},}
  {\protect\JournalTitle{Journal of Lightwave Technology}} \textbf{32},
  1307--1316 (2014).

\bibitem{Liu:18}
H.~Liu, G.~Gredat, G.~Baili, F.~Gutty, F.~Goldfarb, I.~Sagnes, and
  F.~Bretenaker, \enquote{Noise investigation of a dual-frequency {VECSEL} for
  application to cesium clocks,} {\protect\JournalTitle{J. Lightwave Technol.}}
  \textbf{36}, 3882--3891 (2018).

\bibitem{Dumont:14}
P.~Dumont, F.~Camargo, J.~M.~. Danet, D.~Holleville, S.~Guerandel, G.~Pillet,
  G.~Baili, L.~Morvan, D.~Dolfi, I.~Gozhyk, G.~Beaudoin, I.~Sagnes, P.~Georges,
  and G.~Lucas-Leclin, \enquote{Low-noise dual-frequency laser for compact {Cs}
  atomic clocks,} {\protect\JournalTitle{J. Lightwave Technol.}} \textbf{32},
  3817--3823 (2014).

\bibitem{Zanon:2005}
T.~Zanon, S.~Guerandel, E.~de~Clercq, D.~Holleville, N.~Dimarcq, and
  A.~Clairon, \enquote{High contrast {R}amsey fringes with
  coherent-population-trapping pulses in a double lambda atomic system,}
  {\protect\JournalTitle{Phys. Rev. Lett.}} \textbf{94}, 193002--1--193002--4
  (2005).

\bibitem{Liu:18OL}
H.~Liu, G.~Gredat, S.~De, I.~Fsaifes, A.~Ly, R.~Vatr\'{e}, G.~Baili,
  S.~Bouchoule, F.~Goldfarb, and F.~Bretenaker, \enquote{Ultra-low noise
  dual-frequency {VECSEL} at telecom wavelength using fully correlated
  pumping,} {\protect\JournalTitle{Opt. Lett.}} \textbf{43}, 1794--1797 (2018).

\bibitem{Henry:1982}
C.~Henry, \enquote{Theory of the linewidth of semiconductor lasers,}
  {\protect\JournalTitle{IEEE Journal of Quantum Electronics}} \textbf{18},
  259--264 (1982).

\bibitem{Laurain:10}
A.~Laurain, M.~Myara, G.~Beaudoin, I.~Sagnes, and A.~Garnache,
  \enquote{Multiwatt-power highly-coherent compact single-frequency tunable
  vertical-external-cavity-surface-emitting-semiconductor-laser,}
  {\protect\JournalTitle{Opt. Express}} \textbf{18}, 14627--14636 (2010).

\bibitem{Reichling:1994}
M.~Reichling and H.~Gr\"onbeck, \enquote{Harmonic heat flow in isotropic
  layered systems and its use for thin film thermal conductivity measurements,}
  {\protect\JournalTitle{Journal of Applied Physics}} \textbf{75}, 1914--1922
  (1994).

\bibitem{Tricot:2018}
F.~Tricot, \enquote{Analysis and reduction of the frequency instability noise
  sources in a compact {CPT} clock,} Ph.D. thesis, Sorbonne universit\'e
  (2018).

\bibitem{Dumont:15}
P.~Dumont, J.~Danet, F.~Camargo, D.~Holleville, S.~Guerandel, G.~Baili,
  L.~Morvan, G.~Pillet, D.~Dolfi, I.~Gozhyk, G.~Beaudoin, I.~Sagnes,
  P.~Georges, and G.~Lucas-Leclin, \enquote{Evaluation of the noise properties
  of a dual-frequency {VECSEL} for compact cs atomic clocks,}
  {\protect\JournalTitle{Proc.SPIE}} \textbf{9349}, 9349--1 -- 9349--10 (2015).

\bibitem{Francois:14}
B.~Fran\c{c}ois, C.~Calosso, J.~M. Danet, and R.~Boudot, \enquote{A low phase
  noise microwave frequency synthesis for a high-performance cesium vapor cell
  atomic clock,} {\protect\JournalTitle{Review of Scientific Instruments}}
  \textbf{85}, 094709--1--094609--7 (2014).

\bibitem{Camargo:13}
F.~A. Camargo, N.~Girard, J.~M. Danet, G.~Baili, L.~Morvan, D.~Dolfi,
  D.~Holleville, S.~Gu\'{e}randel, I.~Sagnes, P.~Georges, and G.~Lucas-Leclin,
  \enquote{Tunable high-purity microwave signal generation from a
  dual-frequency {VECSEL} at 852 nm,} {\protect\JournalTitle{Proc.SPIE}}
  \textbf{8606}, 8606S--1 -- 8606S--9 (2013).

\bibitem{Chomet:18}
B.~Chomet, J.~Zhao, L.~Ferrieres, M.~Myara, G.~Guiraud, G.~Beaudoin, V.~Lecocq,
  I.~Sagnes, N.~Traynor, G.~Santarelli, S.~Denet, and A.~Garnache,
  \enquote{High-power tunable low-noise coherent source at 1.06 $\mu$m based on
  a surface-emitting semiconductor laser,} {\protect\JournalTitle{Appl. Opt.}}
  \textbf{57}, 5224--5229 (2018).

\bibitem{Link:2018}
S.~M. Link, D.~J. H.~C. Maas, D.~Waldburger, and U.~Keller, \enquote{Dual-comb
  spectroscopy of water vapor with a free-running semiconductor disk laser,}
  {\protect\JournalTitle{Science}} \textbf{356}, 1164--1168 (2017).

\end{thebibliography}

\end{document}